\documentclass[11pt]{article}
\usepackage{amsmath}
\usepackage{amsfonts}
\usepackage{amssymb}
\usepackage{graphicx}
\usepackage{bm}
\usepackage{color}
\usepackage{amscd}

\def\bea{\begin{eqnarray}}
\def\eea{\end{eqnarray}}

\begin{document}
\begin{center}
\LARGE {\bf Entropy Formula and Conserved Charges of Spin-3 Chern-Simons-Like Theories of Gravity }
\end{center}

\begin{center}
{M. R. Setare \footnote{E-mail: rezakord@ipm.ir}\hspace{1mm} ,
H. Adami \footnote{E-mail: hamed.adami@yahoo.com}\hspace{1.5mm} \\
{\small {\em  Department of Science, University of Kurdistan, Sanandaj, Iran.}}}\\

\end{center}

\begin{center}
{\bf{Abstract}}\\
In this paper we present the generalization of Chern-Simons-like theories of gravity (CSLTG) to spin-3. We propose a Lagrangian describing the spin-3 fields coupled to Chern-Simons-like theories of gravity. Then we obtain conserved charges of these theories by using a quasi-local formalism. We find a general formula for entropy of black holes solutions of Spin-3 CSLTG. As an example, we apply our formalism to the spin-3 Generalized minimal massive gravity (GMMG) model. We analysis this model at linearized level and show that this model propagate two massive spin-2 modes and two massive spin-3 modes. We find no-ghost and no-tachyon conditions, which can be satisfy in the parameter space of the model. Then we find energy, angular momentum and entropy of a special black hole solution of this model.
\end{center}

\section{Introduction}
We know that 3-dimensional gravity is the simplest model for studying gravitational dynamics. However since it has rich physics in both classical \cite{t} and quantum versions \cite{w,c}. So there are many motivations for studying gravity in 3 dimensions. By this study we can address conceptual issues of quantum gravity, investigate black hole evaporation, information loss, and black hole microstate counting. Also we can understand the black hole holography deeper. Gauge gravity duality can be extend to the  beyond standard AdS/CFT \cite{k}, such as warped AdS, asymptotic Schrodinger/Lifshitz, non-relativistic CFTs, flat space holography, logarithmic CFTs, and higher spin gravity, which last topic is the subject of this paper.\\
It is well known that Einstein-Hilbert action in the presence of negative cosmological constant in 3-dimension can be reformulated as a Chern-Simons theory with gauge group $SO(2,2)\sim SL(2,\mathbb{R}) \times SL(2,\mathbb{R})$ \cite{a,w}. Similarly, a  $SL(3,\mathbb{R}) \times SL(3,\mathbb{R})$ Chern-Simons theory with the following action describes a three dimensional spin-3 gravity theory \cite{2',2''} ,
\begin{equation}\label{280}
  S_{EH}= S_{CS} [A ^{+}] - S_{CS}[A^{-}],
\end{equation}
where
\begin{equation}\label{290}
   S_{CS} [A] = \frac{l}{8 \pi G} \int tr \left\{ A \wedge d A + \frac{2}{3} A \wedge A \wedge A \right\} ,
\end{equation}
where $l^2>0$ corresponds to a negative cosmological constant, and $G$ is Newton's constant. In the other hand, there is a class of gravitational theories in $(2+1)$-dimension (e.g. Topological massive gravity (TMG) \cite{1'}, New massive gravity (NMG)\cite{6}, Minimal massive gravity (MMG) \cite{7}, Generalized minimal massive gravity (GMMG) \cite{8}, etc), called the Chern-Simons-like theories of gravity \cite{5}. \footnote{Recently, further clarifications on models in 3D are also collected in the book \cite{b3}.} The authors of \cite{9} have done the generalization of topologically massive gravity to
higher spins, specifically spin-3 (see also \cite{10}). In this paper we propose an Lagrangian describing the spin-3 fields coupled to Chern-Simons-like theories of gravity, then we obtain conserved charges of black hole solutions of these theories by using a quasi-local formalism.
In previous paper we have obtained conserved charges of spin-3 topologically massive gravity by this method \cite{4}. In ref. \cite{31} the energy of the higher spin black hole solutions of ordinary higher spin gravity has obtained by canonical formalism (see also \cite{32}).\\
 The authors of \cite{12} have obtained the quasi-local conserved charges for  black holes in any diffeomorphically invariant theory of gravity. By considering an appropriate variation of the metric, they have established a one-to-one correspondence between the ADT approach and the linear Noether expressions. They have extended this work to a theory of gravity containing a gravitational Chern-Simons term in \cite{13}, and have computed the off-shell potential and quasi-local conserved charges of some black holes in TMG.\\
 Our paper is organized as follows. In section 2 we summarize some relevant aspects of spin-3 gravity in three dimensions in the first order formalism. In section 3 we introduce our Lagrangian for spin-3 fields coupled to Chern-Simons-like theories of gravity, which is an extension of the ordinary Chern-Simons-like theories of gravity. One can obtain the spin-3 TMG as a special case of our generic Lagrangian. In section 4 we find the conserved charges of the spin-3 Chern-Simons-like theories of gravity by quasi-local formalism. By using this formalism, one can obtain conserve charges of solutions that are not asymptotically (A)dS. Then in section 5, we consider a black hole solution of the Spin-3 CSLTG. After that, using general formula we obtained for conserved charges, we find a general formula for entropy of black holes solutions of Spin-3 CSLTG. In section 6 we consider spin-3 generalized minimal massive gravity (Spin-3 GMMG) as an example of the Spin-3 CSLTG. In subsection 6.1 we obtain a class of solutions for this model. In subsection 6.2, we do a linear analysis of spin-3 GMMG. We show that this model has two massive spin-2 modes and two massive spin-3 modes. We will obtain the quadratic Lagrangian for the fluctuations about AdS$_{3}$ vacuum with vanishing spin-3 field. Then from quadratic Lagrangian $L^{(2)}$, we find the no ghost conditions for spin-2 and spin-3 modes. Also, we will show that no-tachyon condition is $\left| l m_{I} \right| \geq 1$.  Then in section 7 we find energy, angular momentum and entropy of a special black hole solution of Spin-3 GMMG. In this case, if we set $e_{\mu} ^{\hspace{1.5 mm} ab}= \omega_{\mu} ^{\hspace{1.5 mm} ab}=0 $,  our entropy formula reduces to the entropy formula in the ordinary GMMG which is obtained in the paper \cite{90}. Also in the limiting case $\alpha =0$ , $m^{2} \rightarrow \infty$, where Spin-3 GMMG model reduce to the spin-3 TMG model, our results for energy, angular momentum and entropy reduce to the corresponding results in the spin-3 TMG \cite{4}. Section 8 is devoted to conclusions and discussions.
\section{ Spin-3 gravity in three dimensions}
In this section, we summarize some relevant aspects of spin-3 gravity in three dimensions in the first order formalism. In this frame work, spin-3 gravity can be describes by generalized dreibein and generalized spin-connection and they which take values in the Lie algebra $sl(3,\mathbb{R})$ as follows, respectively \cite{2,3}:
\begin{equation}\label{1}
  e = e _{\mu} \hspace{0.1 mm} ^{A} J _{A} d x ^{\mu}=(e_{\mu} \hspace{0.1 mm} ^{a} J _{a} + e_{\mu} \hspace{0.1 mm} ^{ab} T _{ab})dx ^{\mu},
\end{equation}
\begin{equation}\label{1'}
   \omega = \omega _{\mu} \hspace{0.1 mm} ^{A} J _{A} d x ^{\mu}=(\omega _{\mu} \hspace{0.1 mm} ^{a} J _{a} + \omega _{\mu} \hspace{0.1 mm} ^{ab} T_{ab})d x ^{\mu} _{ab},
\end{equation}
 where $J _{A}$ denotes the generators of $sl(3,\mathbb{R})$ algebra, $A=1, \dots , 8$, $a,b=1,2,3$, and $T_{ab}$ are symmetric and trace-less in the Lorentz indices. The algebra $sl(3,\mathbb{R})$ have eight generators so in the above equations $A=1, \dots , 8$ (Appendices contain more technical details). Translating the frame-like formalism to the metric-like formalism given by the following transformations \cite{1}
\begin{equation}\label{2}
g_{\mu \nu}=\frac{1}{2!} tr(e_{(\mu} e_{\nu )}),
\end{equation}
\begin{equation}\label{2'}
  \varphi _{\mu \nu \lambda} = \frac{1}{3!} tr(e_{(\mu} e_{\nu} e_{\lambda )}).
\end{equation}
The spacetime metric $g_{\mu \nu}$ and the spin-3 field $\varphi _{\mu \nu \lambda}$ both are invariant under the following Lorentz-like gauge transformation
\begin{equation}\label{3}
  \tilde{e}_{\mu} = L e _{\mu} L^{-1},
\end{equation}
where $ L \in SL(3, \mathbb{R})$ and we can write $L = \exp{ (\lambda)}$, where $\lambda$ is the generator of Lorentz-like transformation and it is a $sl(3, \mathbb{R})$ Lie algebra valued quantity,
\begin{equation}\label{4}
 \lambda = \lambda ^{A} J_{A}=\lambda^{a}J_{a}+\lambda^{ab}T_{ab}.
\end{equation}
The exterior covariant derivative define as
\begin{equation}\label{5}
  D_{\mu} e_{\nu} = \partial _{\mu} e_{\nu} + [ \omega _{\mu} , e _{\nu}]  ,
\end{equation}
and under Lorentz-like transformations the generalized spin-connection transforms as
\begin{equation}\label{6}
  \tilde{ \omega } = L \omega L^{-1} +L d L ^{-1},
\end{equation}
where $d$ denotes the ordinary exterior derivative. One can define the total derivative as follows:
\begin{equation}\label{7}
  D ^{(T)} _{\mu} e_{\nu} = \partial _{\mu} e_{\nu} + [ \omega _{\mu} , e _{\nu}] - \Gamma ^{\lambda} _{\mu \nu} e_{\lambda} ,
\end{equation}
where $ \Gamma ^{\lambda} _{\mu \nu} $ is Affine connection. The metric-connection compatibility condition $\nabla _{\lambda} g_{\mu \nu} =0$ leads to the $D ^{(T)} _{\mu} e_{\nu}=0$. Generalized torsion 2-form and generalized curvature 2-form are defined as
\begin{equation}\label{8}
  \mathcal{T}= e_{\lambda} \Gamma ^{\lambda} _{\mu \nu} dx^{\mu} \wedge dx^{\nu} = D e,
\end{equation}
\begin{equation}\label{9}
  \mathcal{R} = d \omega + \omega \wedge \omega \hspace{0.5 cm} ,
\end{equation}
respectively. Also, the ordinary Lie derivative of dreibein along a curve generated by the vector filed $\xi$, $ \pounds _{\xi} e= i_{\xi} d e + d i _{\xi}e$, \footnote{Where $i_{\xi}$ denotes interior product in $\xi$}. can be generalized so that it becomes covariant under Lorentz-like transformations
\begin{equation}\label{10}
  \mathfrak{L}_{\xi} e = \pounds _{\xi} e + [\lambda _{\xi} , e],
\end{equation}
for this purpose, the generator of Lorentz-like gauge transformation $\lambda _{\xi}$ must transforms as
\begin{equation}\label{11}
  \tilde{\lambda _{\xi} } = L \lambda _{\xi} L^{-1} + L \pounds _{\xi} L ^{-1},
\end{equation}
under Lorentz-gauge transformation. In this way, under generalized local translations we have
\begin{equation}\label{12}
  \delta _{\xi} e = \mathfrak{L}_{\xi} e,
\end{equation}
\begin{equation}\label{13}
  \delta _{\xi} \omega = \mathfrak{L}_{\xi} \omega - d \lambda _{\xi},
\end{equation}
and this two equations are covariant under the Lorentz-like gauge transformations as well as diffeomorphisms. \\
We have introduced $\lambda$ to be the generator of Lorentz-like gauge transformation (See Eq.\eqref{4}). In the equation \eqref{3}, $\lambda$ is just an arbitrary function of coordinates. It is easy to see that the ordinary Lie derivative of a Lorentz-like invariant quantity, say $e$, is not Lorentz-like covariant. On the other hand, the change of $e$ under Lorentz-like gauge transformation is given by Eq.\eqref{12}. It was shown that by combining the change due to infinitesimal Lorentz-like gauge transformation with ordinary Lie derivative one can define generalized Lie derivative \eqref{10} which is Lorentz-like covariant provided that $\lambda$ transforms as \eqref{11} under Lorentz-like gauge transformation. Simply one can see form Eq.\eqref{11} that the change of $\lambda$ , under infinitesimal Lorentz-like gauge transformation, is given by $\delta \lambda = - \pounds _{\xi} \lambda$. So, we expect that $\lambda$ to be also a function of $\xi$, i.e. $\lambda =  \lambda _{\xi} (x)$. Another reason for this comes from the fact that we set total variation due to $\xi$ equal to generalized Lie derivative with respect to $\xi$ and $\lambda _{\xi}$ is obliged to generate variation due to $\xi$ (See Eq.\eqref{12}). Thus, in the formalism presented in this paper, $\lambda$ should be a function of coordinates and of $\xi$.\\
The following action describes a three dimensional spin-3 gravity theory
\begin{equation}\label{14}
  S_{EH}= \frac{1}{16 \pi } \int tr \left\{ e \wedge \mathcal{R} +\frac{1}{3l^{2}} e \wedge e \wedge e \right\}.
\end{equation}
The field equations come from above action are
\begin{equation}\label{15}
  \mathcal{T}(\Omega)=0, \hspace{1.5 cm} \mathcal{R}(\Omega)+ \frac{1}{l^{2}} e \wedge e =0,
\end{equation}
where $\omega=\Omega$ is torsion-free spin-connection.\footnote{Torsional extension of Einstein's GR in 3D have been considered in \cite{b1}, also rotating black hole solutions in a generalized topological 3D gravity  with torsion, have analyzed in \cite{b2}.}

\section{Spin-3 Chern-Simons-like theories of gravity}
The ordinary Chern-Simons-like theories of gravity are investigated in some papers, for instance see \cite{5,6,7,8}. These type of theories are an extension of general relativity in 3D which is a gauge theory ( to know more about gauge-theoretic approach to gravity see \cite{130}). The ordinary Chern-Simons-like theories of gravity can be generalize to spin-3 one (Spin-3 CSLTG) by the following Lagrangian
\begin{equation}\label{16}
  L = tr \{ \frac{1}{2} \tilde{g}_{rs} a^{r} \wedge d a^{s} + \frac{1}{3} \tilde{f}_{rst} a^{r} \wedge a^{s} \wedge a^{t} \},
\end{equation}
where $ a^{r} = a^{rA}_{\hspace{3 mm} \mu} J_{A} dx^{\mu} $ are $sl(3,\mathbb{R})$ Lie algebra valued one-forms and $r=1,...,N$ refers to flavour index. Also, $\tilde{g} _{rs}$ is a symmetric constant metric on the flavour space and $\tilde{f} _{rst}$ is a totally symmetric "flavour tensor" which is interpreted as the coupling constant. We take $ a^{r} =\{ e, \omega , h , \cdots \} $, where $ h $ is an auxiliary field and so on. For all of our interesting spin-3 Chernn-Simons-like theories of gravity we have $\tilde{f} _{\omega rs} = \tilde{g}_{rs} $. If one sets $ a^{rab}_{\mu}=0$ and uses Eq.\eqref{114} the Lagrangian \eqref{16} will be reduced to the Lagrangian of ordinary Chern-Simons-like theories of gravity.\\
The arbitrary variation of this Lagrangian given by
\begin{equation}\label{17}
  \delta L = tr \{ \delta a^{r} \wedge E_{r} \} + d \Theta (a, \delta a),
\end{equation}
where
\begin{equation}\label{180}
  E_{r} = \tilde{g}_{rs} d a^{s} + \tilde{f}_{rst} a^{s} \wedge a^{t},
\end{equation}
and
\begin{equation}\label{19}
  \Theta (a, \delta a) = tr \{ \frac{1}{2} \tilde{g}_{rs} \delta a^{r} \wedge  a^{s} \}.
\end{equation}
The equations of motion of these theories are $E_{r}=0$ and $\Theta (a, \delta a)$ is surface term. By considering equations \eqref{12} and \eqref{13}, under generalized local translations, $a^{r}$ transforms as
\begin{equation}\label{20}
  \delta _{\xi} a^{r} = \mathfrak{L}_{\xi} a^{r} - \delta ^{r}_{\omega} d \lambda _{\xi},
\end{equation}
where $ \delta ^{r}_{\omega} $ Kronecker delta. If we take the non-zero components of the flavour metric and the flavour tensor as
\begin{equation}\label{21}
\begin{split}
     & \tilde{g} _{e \omega} = - \sigma , \hspace{0.5 cm} \tilde{g} _{eh} = 1 , \hspace{0.5 cm} \tilde{g} _{\omega \omega} = \frac{1}{\mu} , \\
     & \tilde{f} _{e \omega \omega}= -\sigma , \hspace{0.5 cm} \tilde{f} _{e h \omega}=1 , \hspace{0.5 cm} \tilde{f} _{\omega \omega \omega} = \frac{1}{\mu} ,  \hspace{0.5 cm} \tilde{f} _{eee}=\Lambda ,
\end{split}
\end{equation}
the Spin-3 CSLTG reduce to spin-3 topologically massive gravity (Spin-3 TMG) which is investigated at the linearized level in the papers \cite{9,10}. In paper \cite{4} we have obtained conserved charges of this model.
\section{Quasi-local conserved charges}
In this section we want to find the conserved charges of the spin-3 Chern-Simons-like theories of gravity. Since $e$, $\omega$ and auxiliary fields appeared in CSLTG are invariant under general coordinates transformation it is clear that the Lagrangian of CSLTG \eqref{16} is invariant under general coordinates transformation. Now we recall that under general Lorentz-like gauge transformation $e$ and $\omega$ transform as \eqref{3} and \eqref{6}, respectively. Also it is easy to check that under general Lorentz-like gauge transformation generalized curvature 2-form \eqref{9} and generalized torsion 2-form \eqref{8} transform as $\tilde{\mathcal{R}}=L\mathcal{R}L^{-1}$ and $\tilde{\mathcal{T}}=L\mathcal{T}L^{-1}$. Therefore it is inferred that three dimensional spin-3 gravity theory described by action \eqref{14} is a covariant theory under general Lorentz-like gauge transformation. Also, it is clear that equations of motion of spin-3 gravity theory, Eq.\eqref{15}, are covariant under general Lorentz-like gauge transformation. Nevertheless, there may be exist theories which are not covariant under general Lorentz-like gauge transformation. In other words, theories described by the Lagrangian \eqref{16} may contain terms that can break covariance under general Lorentz-like gauge transformation. Spin-3 Topologically Massive Gravity term, $\frac{1}{2 \mu} tr\left( \omega \wedge d \omega + \frac{2}{3} \omega\wedge \omega \wedge \omega  \right)$, is an example of such terms. This term is appeared in the spin-3 Topologically Massive Gravity \cite{9,10,4} and in spin-3 generalized minimal massive gravity which will be introduced in section 6.
Under generalized local translations, the Lagrangian \eqref{16} transforms as
\begin{equation}\label{18}
  \delta _{\xi} L = \mathfrak{L}_{\xi} L + d \psi _{\xi},
\end{equation}
where
\begin{equation}\label{23}
  \psi _{\xi} = tr \{ \frac{1}{2} \tilde{g} _{\omega r} d \lambda _{\xi} \wedge a^{r} \}.
\end{equation}
which is equivalent to the statement that a symmetry is a transformation which leaves the Lagrangian form invariant, up to a total derivative. Despite the fact that a Lagrangian is not invariant under general Lorentz-like gauge transformation, if a Lagrangian behaves like \eqref{18} under generalized local translations, then $\xi$ could be a symmetry generator. Although the Lagrangian \eqref{16} is not invariant under general Lorentz-like gauge transformation, but by virtue of equations \eqref{18} and \eqref{23}, it is invariant under the infinitesimal Lorentz-like gauge transformation. Also, it is enough in obtaining generally covariant equations of motion that Lagrangian behaves like \eqref{18} under generalized local translations.
 Now, we consider the variation of Lagrangian with respect to generalized local translations
\begin{equation}\label{20'}
  \delta _{\xi} L = tr \{ \delta _{\xi} a^{r} \wedge E_{r} \} + d \Theta (a, \delta _{\xi} a),
\end{equation}
Comparing Eq.\eqref{18} with Eq.\eqref{20'} leads to the following relation
\begin{equation}\label{210}
  d \left( \Theta (a, \delta _{\xi} a) - i_{\xi} L - \psi _{\xi} \right) = - tr \{ \delta _{\xi} a^{r} \wedge E_{r} \}.
\end{equation}
 We can rewrite Eq.\eqref{20} as
\begin{equation}\label{26}
  \begin{split}
       & \delta _{\xi} a^{r^{\prime}} = D i_{\xi} a^{r^{\prime}} + i_{\xi} D a^{r^{\prime}} + [ (\lambda _{\xi} - i _{\xi} \omega ), a^{r^{\prime}} ], \\
       & \delta _{\xi} \omega = i_{\xi} \mathcal{R} - D ( \lambda _{\xi}- i_{\xi} \omega),
  \end{split}
\end{equation}
where the prime on $r $ indicates that the sum run over all the flavour indices except $\omega$. Now, we substitute equations \eqref{26} into Eq.\eqref{21} and we find that
\begin{equation}\label{27}
  \begin{split}
   d \mathfrak{J} _{\xi} = & - tr \{ (\lambda _{\xi} - i _{\xi} \omega ) (D E_{\omega} + a^{r^{\prime}} \wedge E _{r^{\prime}} - E_{r^{\prime}}  \wedge a^{r^{\prime}} ) \} \\
     & + tr \{ i_{\xi} a^{r^{\prime}} D E_{r^{\prime}} - i_{\xi} D a^{r^{\prime}} \wedge E_{r^{\prime}} - i_{\xi} \mathcal{R} \wedge E_{\omega} \},
\end{split}
\end{equation}
where $\mathfrak{J} _{\xi}$ is given by
\begin{equation}\label{28}
  \mathfrak{J} _{\xi} = \Theta (a , \delta _{\xi} a) - i_{\xi} L - \psi _{\xi} + tr \{ i_{\xi} a^{r} E_{r} - \lambda _{\xi} E _{\omega} \}.
\end{equation}
 We expect that the last line in Eq.\eqref{27} can be rewritten as
\begin{equation}\label{25}
  tr \{ i_{\xi} a^{r^{\prime}} D E_{r^{\prime}} - i_{\xi} D a^{r^{\prime}} \wedge E_{r^{\prime}} - i_{\xi} \mathcal{R} \wedge E_{\omega} \} =tr \{ i_{\xi} a^{r^{\prime}} X_{r^{\prime}}(a) \},
\end{equation}
see \cite{4} for the Spin-3 TMG case. In order to have an off-shell conserved current $\mathfrak{J} _{\xi}$ we must have
\begin{equation}\label{30}
  D E_{\omega} + a^{r^{\prime}} \wedge E _{r^{\prime}} - E_{r^{\prime}}  \wedge a^{r^{\prime}} =0 , \hspace{1 cm} X_{r^{\prime}}(a)=0.
\end{equation}
These equations give us the Bianchi identities and they reduce to the following one for the Spin-3 TMG model \cite{4}
\begin{equation}\label{31}
\begin{split}
     & D \mathcal{R} = 0, \\
     & D \mathcal{T} + e \wedge \mathcal{R} - \mathcal{R} \wedge e =0.
\end{split}
\end{equation}
In this way, through the Bianchi identities, $\mathfrak{J} _{\xi}$ is conserved off-shell, i.e. $ d \mathfrak{J} _{\xi} = 0$. Accordingly, by virtue of Poincare lemma, we find that
\begin{equation}\label{32}
  \mathfrak{J} _{\xi} = d K _{\xi},
\end{equation}
where $K_{\xi}$ is given by
\begin{equation}\label{33}
  K_{\xi} = tr \{ \frac{1}{2} \tilde{g} _{rs} i_{\xi} a^{r} a^{s} - \tilde{g} _{r \omega} \lambda _{\xi} a^{r} \}.
  \end{equation}
It is straightforward to show that under generalized local translations, $\Theta (a ,\delta a)$ transforms as
\begin{equation}\label{34}
 \delta _{\xi} \Theta (a ,\delta a) = \mathfrak{L}_{\xi} \Theta (a ,\delta a) + \Pi _{\xi},
\end{equation}
where
\begin{equation}\label{35}
  \Pi _{\xi} = tr \{ \frac{1}{2} \tilde{g} _{r \omega} d \lambda _{\xi} \wedge \delta a ^{r} \}.
\end{equation}
By taking arbitrary variation from the equation \eqref{32} and using Eq.\eqref{28} we have
\begin{equation}\label{36}
\begin{split}
   d \left( \delta K _{\xi} - i _{\xi} \Theta (a ,\delta a) \right) =& tr \{ \delta a ^{r} \wedge i_{\xi} E _{r} + i_{\xi} a^{r} \delta E _{r} - \lambda _{\xi} \delta E _{\omega} \} \\
     & + \delta \Theta (a ,\delta _{\xi} a) - \delta _{\xi} \Theta (a ,\delta a),
\end{split}
\end{equation}
where we have used the relation $\Pi _{\xi} = \delta \psi _{\xi}$ in the last step of calculation. On the other hand, if we demand that $\xi$ be a Killing vector field admitted by the whole of spacetime, we have the following configuration space result given in \cite{11}
\begin{equation}\label{37}
  \delta \Theta (a , \delta _{\xi} a) - \delta _{\xi} \Theta (a , \delta a)=0,
\end{equation}
then the right hand side of the equation \eqref{36} is the off-shell ADT current
\begin{equation}\label{38}
  \mathcal{J} _{ADT} =  tr \{ \delta a ^{r} \wedge i_{\xi} E _{r} + i_{\xi} a^{r} \delta E _{r} - \lambda _{\xi} \delta E _{\omega} \}.
\end{equation}
By substituting the components of the flavor metric and flavour tensor from Eq.\eqref{21} into above equation, current $\mathcal{J} _{ADT}$ reduces to the off-shell ADT current appearing in the Spin-3 TMG\cite{4}. By demanding the field equations $ E _{r} =0 $, and the linearized field equations $ \delta E _{r} =0 $, the right hand side of the equation \eqref{36} is simply the symplectic current
\begin{equation}\label{39}
  \Omega _{symplectic} (a , \delta a , \delta _{\xi} a) = \delta \Theta (a , \delta _{\xi} a) - \delta _{\xi} \Theta (a , \delta a).
\end{equation}
Hence, it seems reasonable to generalize the off-shell ADT current so that it becomes conserved when spacetime admits $\xi$ as an asymptotically Killing vector field rather than a Killing vector field admitted by the whole of spacetime. So, we can define generalized off-shell ADT current as
\begin{equation}\label{40}
  \tilde{\mathcal{J}} _{ADT} = \mathcal{J} _{ADT} + \Omega _{symplectic}.
\end{equation}
As mentioned above, this generalized off-shell ADT current reduces to the ordinary one when $\xi$ is a Killing vector field admitted by the whole of spacetime and it reduces to the symplectic current when $ E _{r} = \delta E _{r} =0 $, and $\xi$ is an asymptotically Killing vector field. By substituting Eq.\eqref{40} into \eqref{36} we have
\begin{equation}\label{41}
   \tilde{\mathcal{J}} _{ADT} = d \left( \delta K _{\xi} - i _{\xi} \Theta (a ,\delta a) \right),
\end{equation}
so, it is obvious that the generalized off-shell ADT current is conserved for Killing vectors which are admitted by the whole of spacetime and asymptotically Killing vectors. Hence we can define generalized off-shell ADT conserved charge as
\begin{equation}\label{42}
  \tilde{\mathcal{Q}} _{ADT} (a, \delta a ; \xi)= \delta K _{\xi} - i _{\xi} \Theta (a ,\delta a),
\end{equation}
for which $\tilde{\mathcal{J}} _{ADT} = d \tilde{\mathcal{Q}} _{ADT}$. Now, in the manner of papers \cite{12,13}, we define quasi-local conserved charge corresponds to an asymptotically Killing vector field $\xi$ as
\begin{equation}\label{43}
  Q(\xi) =\frac{1}{8 \pi } \int_{0}^{1} ds \int _{\Sigma} \tilde{\mathcal{Q}} _{ADT} (a | s) ,
\end{equation}
where $\Sigma$ denotes an arbitrary codimension two space-like surface and integration with respect to $s$ runs over an one-parameter path in the solution space, where $s = 0$ and $s = 1$ are correspond to the background solution and the interested solution, respectively.\\
By substituting the explicit forms of $K_{\xi}$ and $\Theta (a ,\delta a)$ into Eq.\eqref{43} we find that
\begin{equation}\label{44}
   Q(\xi) =\frac{1}{8 \pi } \int_{0}^{1} ds \int _{\Sigma} tr \{ ( \tilde{g} _{rs} i_{\xi} a^{s} - \tilde{g} _{r \omega} \lambda _{\xi} ) \delta a^{r} \}.
\end{equation}
The advantage of this formalism is that it is applicable to solutions that are not asymptotically (A)dS.
 Assuming that $ (\delta _{1} \delta _{2} - \delta _{2} \delta _{1}) a^{r} =0$, it is clear that
 the conserved charge \eqref{44} has to satisfy the integrability condition
$ (\delta _{1} \delta _{2} - \delta _{2} \delta _{1}) Q(\xi) =0$ \cite{11}.
 Hence, the result of Eq.\eqref{44} does not depend on the given path on the solution space.

\section{A general formula for the entropy of black holes}
Let us consider a black hole solution of the Spin-3 CSLTG. We take the codimension two surface $\Sigma$ to be the bifurcate surface $\mathcal{B} $. Suppose that $\xi$ is the Killing vector field which generates the Killing horizon, so we must set $\xi=0$ on $\mathcal{B} $.\footnote{ Here, we consider stationary black hole solutions and we assume that the event horizon of considered black hole is a non-degenerate Killing horizon. As we know, a cross-section of non-degenerate Killing horizon is bifurcate surface where $\xi=0$.} Since Eq.\eqref{44} is conserved for Killing vectors which are admitted by the whole of spacetime and for asymptotically Killing vectors, so we can use Eq.\eqref{44} to find the entropy of black holes in the context of considered theory as a conserved charge corresponds to the Killing vector field which generates the Killing horizon. Thus, the conserved charge expression \eqref{44} reduces to
\begin{equation}\label{45}
   Q(\xi) = - \frac{1}{8 \pi } \int_{0}^{1} ds \int _{\mathcal{B}} tr \{ \tilde{g} _{r \omega} \lambda _{\xi} \delta a^{r} \}.
\end{equation}
Now, we take $s=0$ and $s=1$ correspond to the considered black hole space-time and the perturbed one respectively. Thus, the equation \eqref{44} becomes
\begin{equation}\label{46}
   \delta Q(\xi) = - \frac{1}{8 \pi } \tilde{g} _{r \omega} \int _{\mathcal{B}} tr \{ \lambda _{\xi} \delta a^{r} \}.
\end{equation}
Since $\lambda _{\xi}$ do not depends on fields at all, we obtain
\begin{equation}\label{47}
   Q(\xi) = - \frac{1}{8 \pi } \tilde{g} _{r \omega} \int _{\mathcal{B}} tr \{ \lambda _{\xi} a^{r} \}.
\end{equation}
By demanding that the generalized Lie derivative of generalized dreibein vanishes explicitly when $\xi$ is a Killing vector
 field, one can find an expression for $\lambda _{\xi}$ \cite{4}
\begin{equation}\label{173}
  \lambda _{\xi}= i _{\xi} \omega + \frac{1}{2} [e ^{\nu} , D_{\nu} (i_{\xi} e) +(i_{\xi}\mathcal{T})_{\nu} ].
\end{equation}
One can use Eq.\eqref{3} and Eq.\eqref{6} to show that Eq.\eqref{173} satisfy the transformation property \eqref{11}.
It was shown that Eq.\eqref{173} reduces to \cite{4}
\begin{equation}\label{48}
  \lambda_{\xi} = \frac{\kappa}{\sqrt{ g_{\phi \phi}}} e _{\phi},
\end{equation}
on bifurcate surface $\mathcal{B}$, where $\kappa$ is surface gravity. By substituting Eq.\eqref{48} into Eq.\eqref{47}, we can define the entropy of considered black hole solution as
\begin{equation}\label{49}
   S= \frac{8 \pi}{\kappa} Q(\xi) = - \tilde{g} _{r \omega} \int _{\mathcal{B}} \frac{d \phi}{\sqrt{ g_{\phi \phi}}} tr \{ e_{\phi} a^{r}_{\phi} \} .
\end{equation}

\section{Spin-3 generalized minimal massive gravity}
In this section, we introduce spin-3 generalized minimal massive gravity then we analysis it at linearized level.
\subsection{The Model}
We consider spin-3 generalized minimal massive gravity (Spin-3 GMMG) as an example of the Spin-3 CSLTG. The ordinary generalized minimal massive gravity have studied originally in \cite{8}. The generalized minimal massive gravity theory is realized
by adding the CS deformation term, the higher derivative deformation term, and an extra term
to pure Einstein gravity with a negative cosmological constant. In \cite{8} it is discussed that this theory is free of negative-energy bulk modes, and also avoids the aforementioned ``bulk-boundary unitarity clash''. By a Hamiltonian analysis one can show that the GMMG model has
no Boulware-Deser ghosts and this model propagate only two physical modes.
In this model, there are four flavours of Lie algebra valued one-form $ a^{r}=\{ e, \omega, h, f \} $ and the non-zero components of the flavour metric and the flavour tensor are
\begin{equation}\label{50}
\begin{split}
     & \tilde{g} _{e \omega} = - \sigma , \hspace{0.5 cm} \tilde{g} _{eh} = 1 , \hspace{0.5 cm} \tilde{g} _{f \omega} = -\frac{1}{m^{2}} , \hspace{0.5 cm} \tilde{g} _{\omega \omega} = \frac{1}{\mu} , \\
     & \tilde{f} _{e \omega \omega}= -\sigma , \hspace{0.5 cm} \tilde{f} _{e h \omega}=1 , \hspace{0.5 cm} \tilde{f} _{f \omega \omega} = -\frac{1}{m^{2}} , \hspace{0.5 cm} \tilde{f} _{\omega \omega \omega} = \frac{1}{\mu}, \\
     & \tilde{f} _{eee}=\Lambda _{0} , \hspace{0.5 cm} \tilde{f} _{ehh}= \alpha , \hspace{0.5 cm} \tilde{f} _{eff}= - \frac{1}{m^{2}} .
\end{split}
\end{equation}
Thus, equations of motion \eqref{180} reduce to
\begin{equation}\label{51}
  \begin{split}
       & -\sigma \mathcal{R} (\omega) + \Lambda _{0} e \wedge e + D (\omega) h + \alpha h \wedge h - \frac{1}{m^{2}} f \wedge f =0 , \\
       & -\sigma \mathcal{T} (\omega) + \frac{1}{\mu} \mathcal{R} (\omega) +e \wedge h + h \wedge e - \frac{1}{m^{2}} D (\omega) f =0 , \\
       & \mathcal{R} (\omega)+  e \wedge f + f \wedge e =0 , \\
       & \mathcal{T} (\omega)+ \alpha ( e \wedge h + h \wedge e ) =0.
  \end{split}
\end{equation}
It is clear that the Spin-3 GMMG is not a torsion-free theory, but by redefinition of generalized spin-connection, as $\omega = \Omega - \alpha h $, we make it torsion-free, in this case we have
\begin{equation}\label{52}
\begin{split}
     & \mathcal{R} (\omega) = \mathcal{R} (\Omega) - \alpha D (\Omega) h + \alpha ^{2} h \wedge h , \\
     & D (\omega) f =  D (\Omega) f - \alpha ( h \wedge f + f \wedge h ),
\end{split}
\end{equation}
 then equations of motion \eqref{51} can be rewritten as
\begin{equation}\label{53}
  -\sigma \mathcal{R} (\Omega) + \Lambda _{0} e \wedge e +(1+ \alpha \sigma) ( D (\Omega) h - \alpha h \wedge h ) - \frac{1}{m^{2}} f \wedge f =0 ,
\end{equation}
\begin{equation}\label{54}
  \begin{split}
       & \mathcal{R} (\Omega) - \alpha D (\Omega) h + \alpha ^{2} h \wedge h + \mu (1+ \alpha \sigma ) (e \wedge h + h \wedge e) \\
       & - \frac{\mu}{m^{2}} D (\Omega) f + \frac{\mu \alpha}{m^{2}} (f \wedge h + h \wedge f) =0 ,
  \end{split}
\end{equation}
\begin{equation}\label{55}
  \mathcal{R} (\Omega) - \alpha D (\Omega) h + \alpha ^{2} h \wedge h +  e \wedge f + f \wedge e =0,
\end{equation}
\begin{equation}\label{56}
  \mathcal{T} (\Omega) =0.
\end{equation}
Now, we want to find some solutions of the considered model. For this purpose, we consider the following ansatz for $h$ and $f$
\begin{equation}\label{57}
  h= \beta e , \hspace{1 cm} f = \gamma e,
\end{equation}
where $e$ is generalized dreibein and $\beta$ and $\gamma$ are constants. By substituting Eq.\eqref{57} into equations \eqref{53}-\eqref{55} we have
\begin{equation}\label{58}
  \begin{split}
       & -\sigma \mathcal{R}(\Omega)+ \left( \Lambda _{0} - \alpha (1+ \alpha \sigma) \beta ^{2} - \frac{\gamma ^{2}}{m^{2}} \right) \ e \wedge e =0, \\
       & \mathcal{R}(\Omega)+ \left( \alpha ^{2} \beta ^{2} + 2 \mu (1+\alpha \sigma) \beta + \frac{2 \mu \alpha}{m^{2}} \beta \gamma \right) e \wedge e =0 , \\
       & \mathcal{R}(\Omega)+ \left( \alpha ^{2} \beta ^{2} + 2 \gamma \right) e \wedge e =0.
  \end{split}
\end{equation}
By comparing these equations with Eq.\eqref{15} we find that
\begin{equation}\label{59}
  \Lambda _{0} - \alpha (1+ \alpha \sigma) \beta ^{2} - \frac{\gamma ^{2}}{m^{2}} = -\frac{\sigma}{l^{2}},
\end{equation}
\begin{equation}\label{60}
  \alpha ^{2} \beta ^{2} + 2 \mu (1+\alpha \sigma) \beta + \frac{2 \mu \alpha}{m^{2}} \beta \gamma = \frac{1}{l^{2}},
\end{equation}
\begin{equation}\label{61}
  \alpha ^{2} \beta ^{2} + 2 \gamma = \frac{1}{l^{2}}.
\end{equation}
In this way, all solutions of the spin-3 gravity (for instance, see \cite{14,15,16}) are solutions of the Spin-3 GMMG when $\beta$, $\gamma$ and parameters of the considered model satisfy equations \eqref{59}- \eqref{61}. From equations \eqref{59}-\eqref{61}, one finds that
\begin{equation}\label{133}
  l_{\pm}^{2}=\left[ \alpha^{2} \beta^{2}+ 2 \sigma m^{2} \pm 2 \sqrt{m^{2}\left(\sigma^{2} m^{2} - \alpha \beta ^{2} + \Lambda_{0}\right)}\right]^{-1},
\end{equation}
\begin{equation}\label{134}
  \gamma_{\pm}= \sigma m^{2} \pm \sqrt{m^{2}\left(\sigma^{2} m^{2} - \alpha \beta ^{2} + \Lambda_{0}\right)},
\end{equation}
where $\beta$ should be satisfy following quartic equation
\begin{equation}\label{135}
  \begin{split}
       & - \Lambda_{0} m^{4} + 2\mu m^{2} \left[ \alpha \Lambda_{0} - \sigma m^{2} (1+\alpha\sigma) \right] \beta \\
       & +\left[ \mu^{2} m^{2} (1+\alpha\sigma) (1+3\alpha\sigma) + \alpha \left( m^{4} - \alpha \mu^{2} \Lambda_{0} \right) \right] \beta ^{2}\\
       & -2\mu \alpha^{2} m^{2} \beta ^{3} + \mu \alpha^{3} \beta^{4}=0.
  \end{split}
\end{equation}
We have following conditions on roots of quartic equation \eqref{135}
\begin{equation}\label{136}
  \begin{split}
       & \beta^{2} \leq \alpha^{-1} \left( \sigma^{2} m^{2} +\Lambda_{0} \right) \hspace{0.5 cm} \text{for} \hspace{0.5 cm} \alpha > 0, \\
       & \beta^{2} \geq \alpha^{-1} \left( \sigma^{2} m^{2} +\Lambda_{0} \right) \hspace{0.5 cm} \text{for} \hspace{0.5 cm} \alpha < 0,
  \end{split}
\end{equation}
which ensure that $l^{2}$ and $\gamma$ are real. It should be noted that $\beta$ and $\gamma$ could not be complex numbers because they are appear in the energy, angular momentum and entropy experssions (see equations \eqref{68}, \eqref{71} and \eqref{74}).
\subsection{Linearized Analysis}\label{S.A}
The Lagrangian of spin-3 GMMG up to a surface term can be written as
\begin{equation}\label{100}
\begin{split}
   L_{\text{spin-3 GMMG}}= tr \biggl\{ & - \sigma e \wedge \mathcal{R}(\omega) + \frac{\Lambda _{0}}{3} e \wedge e \wedge e + h \wedge \mathcal{T}(\omega) \\
     & + \alpha e \wedge h \wedge h + \frac{1}{2 \mu} \left( \omega \wedge d \omega + \frac{2}{3} \omega\wedge \omega \wedge \omega  \right) \\
     & - \frac{1}{m^{2}} \left( f \wedge \mathcal{R}(\omega) + e \wedge f \wedge f \right) \biggr\}.
\end{split}
\end{equation}
We suppose that the background generalized dreibein and torsion-free generalized spin-connection are given by $\bar{e}$ and $\bar{\Omega}$ respectively, where $\bar{e} _{\mu} \hspace{0.1 mm} ^{ab} = 0$ and $\bar{\Omega} _{\mu} \hspace{0.1 mm} ^{ab} = 0$.\footnote{Refer to \cite{7,100} to see such analysis for minimal massive gravity, and new version of generalized zwei-dreibein gravity, respectively. See also \cite{9,10} for spin-3 topologically massive gravity.} Also, $\bar{e} _{\mu} \hspace{0.1 mm} ^{a}$ and $\bar{\Omega} _{\mu} \hspace{0.1 mm} ^{a}$ describe AdS$_{3}$ vacuum. Then we can take $\bar{h}=  \beta \bar{e}$ and $\bar{f}=  \gamma \bar{e}$, where $ \beta$ and $\gamma$ are just constant parameters, and these solve equations of motion \eqref{53}-\eqref{56} provided that equations \eqref{59}-\eqref{61} are satisfied. Therefore, we have
\begin{equation}\label{101}
  D(\bar{\Omega}) \bar{e} =0, \hspace{1 cm} \mathcal{R}(\bar{\Omega}) + \frac{1}{l^{2}} \bar{e} \wedge \bar{e}=0.
\end{equation}
We now expand $e$, $\omega$, $h$ and $f$ about the background as follows:
\begin{equation}\label{102}
  \begin{split}
     e= \bar{e} + \hat{\varepsilon} u, \hspace{1.5 cm}& \hspace{1 cm}  \Omega= \bar{\Omega} + \hat{\varepsilon} v \\
     h= \beta ( \bar{e} + \hat{\varepsilon} u) + \hat{\varepsilon} w, & \hspace{1 cm}  f= \gamma( \bar{e} + \hat{\varepsilon} u) + \hat{\varepsilon} z
  \end{split}
\end{equation}
where $\hat{\varepsilon}$ is a small expansion parameter. By substituting these expressions into the Lagrangian \eqref{100}, and using \eqref{59}-\eqref{61} which vanish the linear term in the expansion of the Lagrangian, we find that the quadratic Lagrangian for the fluctuations $u$, $v$, $w$ and $z$ is given by
\begin{equation}\label{103}
  \begin{split}
     L^{(2)} = tr \biggl\{ & \frac{1}{2 \mu l^{2}} u \wedge \bar{D} u + \frac{1}{2 \mu} v \wedge \bar{D} v - \left( \sigma + \frac{\alpha \beta}{\mu} + \frac{\gamma}{m^{2}}\right) u \wedge \bar{D} v + \frac{\alpha ^{2}}{2 \mu} w \wedge \bar{D} w\\
       &  + \left( 1+ \alpha \sigma + \frac{\alpha ^{2} \beta}{\mu} + \frac{\alpha \gamma}{m^{2}}\right) u \wedge \bar{D} w - \frac{\alpha}{\mu} v \wedge \bar{D} w - \frac{1}{m^{2}} v \wedge \bar{D} z\\
       & + \frac{\alpha \beta}{m^{2}} u \wedge \bar{D} z + \frac{\alpha}{m^{2}} z \wedge \bar{D} w - \left( \frac{\sigma}{l^2} + \frac{\alpha \beta}{\mu l^{2}} + \frac{2 \gamma ^{2}}{m^{2}}\right) \bar{e} \wedge u \wedge u\\
       & - \left( \sigma + \frac{\alpha \beta}{\mu} + \frac{\gamma}{m^{2}}\right) \bar{e} \wedge v \wedge v + \frac{1}{\mu l^{2}} \left(  \bar{e} \wedge u \wedge v + \bar{e} \wedge v \wedge u \right) \\
       & - \alpha \left( 1+ \alpha \sigma + \frac{\alpha ^{2} \beta}{\mu} + \frac{\alpha \gamma}{m^{2}}\right) \bar{e} \wedge w \wedge w - \frac{\alpha}{\mu l^{2}} \left(  \bar{e} \wedge u \wedge w + \bar{e} \wedge w \wedge u \right) \\
       & + \left( 1+ \alpha \sigma + \frac{\alpha ^{2} \beta}{\mu} + \frac{\alpha \gamma}{m^{2}}\right) \left(  \bar{e} \wedge v \wedge w + \bar{e} \wedge w \wedge v \right)\\
       & + \frac{\alpha \beta}{m^{2}} \left(  \bar{e} \wedge v \wedge z + \bar{e} \wedge z \wedge v \right) - \frac{1}{m^{2} l^{2}} \left(  \bar{e} \wedge u \wedge z + \bar{e} \wedge z \wedge u \right) \biggr\}
  \end{split}
\end{equation}
where $\bar{D} = D(\bar{\Omega})$. One can extracts the linearized equations of motion from the Lagrangian \eqref{103}, or equivalently, one can uses Eq.\eqref{102} to linearize the equations of motion \eqref{53}-\eqref{56}, and then we will have
\begin{equation}\label{104}
\begin{split}
     & \bar{D} u + \left( \bar{e} \wedge v + v \wedge \bar{e} \right)= 0, \\
     & \bar{D} v + \frac{1}{l^{2}} \left( \bar{e} \wedge u + u \wedge \bar{e} \right) + \left[ 1+\alpha \sigma - \frac{\alpha \gamma}{m^{2}}\right] \left( \bar{e} \wedge z + z \wedge \bar{e} \right) = 0, \\
     & \bar{D} w - \alpha \beta \left( \bar{e} \wedge w + w \wedge \bar{e} \right)+ \left[ \sigma - \frac{\gamma}{m^{2}} \right] \left( \bar{e} \wedge z + z\wedge \bar{e} \right) =0, \\
     & \bar{D} z + \left[ \frac{m^{2}}{\mu} - \alpha \beta \right] \left( \bar{e} \wedge z + z \wedge \bar{e} \right) - m^{2} \left[ 1+\alpha \sigma + \frac{\alpha \gamma}{m^{2}} \right] \left( \bar{e} \wedge w + w \wedge \bar{e} \right) =0.
\end{split}
\end{equation}
Now, we introduce following transformations from $(u,v,w,z)$ to new Lie algebra valued one-form fluctuations $(q_{+},q_{-},q_{1},q_{2})$:
\begin{equation}\label{105}
  \begin{split}
     u =  & B_{+} q_{+} + B_{-} q_{-} + B_{1} q_{1} + B_{2} q_{2}, \\
     v =  & m_{+} B_{+} q_{+} + m_{-} B_{-} q_{-} + m_{1} B_{1} q_{1} + m_{2} B_{2} q_{2}, \\
     w =  & C_{1} B_{1} q_{1} + C_{2} B_{2} q_{2}, \\
     z =  & F_{1} B_{1} q_{1} + F_{2} B_{2} q_{2},
  \end{split}
\end{equation}
where $ (B_{+},B_{-},B_{1},B_{2})$ are arbitrary constants and $(m_{+},m_{-},m_{1},m_{2})$ are given by
\begin{equation}\label{106}
  \begin{split}
       & m _{+} = \frac{1}{l}, \hspace{1 cm} m _{-} = -\frac{1}{l} \\
       & m _{1} = \frac{m^{2}}{2 \mu} - \alpha \beta + \left[ \frac{m^{4}}{4 \mu ^{2}} + \gamma + \frac{\alpha \gamma ^{2}}{m^{2}} - (1+ \alpha \sigma) m^{2} \sigma \right]^{\frac{1}{2}}, \\
       & m _{2} = \frac{m^{2}}{2 \mu} - \alpha \beta - \left[ \frac{m^{4}}{4 \mu ^{2}} + \gamma + \frac{\alpha \gamma ^{2}}{m^{2}} - (1+ \alpha \sigma) m^{2} \sigma \right]^{\frac{1}{2}},
  \end{split}
\end{equation}
also,
\begin{equation}\label{107}
  \begin{split}
     C_{1} = \frac{\left( m_{1}^{2}- \frac{1}{l^{2}} \right) \left( m_{2} + \alpha \beta \right)}{m^{2} \left[ \left(1+\alpha \sigma\right)^{2} - \left( \frac{\alpha \gamma}{m^{2}} \right)^{2}\right]}, & \hspace{0.7 cm} C_{2} = \frac{\left( m_{2}^{2}- \frac{1}{l^{2}} \right) \left( m_{1} + \alpha \beta \right)}{m^{2} \left[ \left(1+\alpha \sigma\right)^{2} - \left( \frac{\alpha \gamma}{m^{2}} \right)^{2}\right]}, \\
      F_{1} = \frac{\left( m_{1}^{2}- \frac{1}{l^{2}} \right)}{\left[ 1+\alpha \sigma -  \frac{\alpha \gamma}{m^{2}} \right]}, &\hspace{0.7 cm} F_{2} = \frac{\left( m_{2}^{2}- \frac{1}{l^{2}} \right)}{\left[ 1+\alpha \sigma -  \frac{\alpha \gamma}{m^{2}} \right]}.
  \end{split}
\end{equation}
By using the transformations \eqref{105} we can diagonalize the linearized equations \eqref{104} as follows
\begin{equation}\label{108}
  \begin{split}
       & \bar{D} q_{+} + m_{+} \left( \bar{e} \wedge q_{+} +  q_{+} \wedge \bar{e} \right)=0, \\
       & \bar{D} q_{-} + m_{-} \left( \bar{e} \wedge q_{-} +  q_{-} \wedge \bar{e} \right)=0, \\
       & \bar{D} q_{1} + m_{1} \left( \bar{e} \wedge q_{1} +  q_{1} \wedge \bar{e} \right)=0, \\
       & \bar{D} q_{2} + m_{2} \left( \bar{e} \wedge q_{2} +  q_{2} \wedge \bar{e} \right)=0.
  \end{split}
\end{equation}
In this manner, we can expect that the transformations \eqref{105} diagonalize the Lagrangian \eqref{103}. So we can rewrite the Lagrangian \eqref{103} in the diagonalized form, in terms of new Lie algebra valued 1-form fields,
\begin{equation}\label{109}
  \begin{split}
     - L^{(2)} = tr \biggl\{ & \left[ \sigma + \frac{\alpha \beta}{\mu} + \frac{\gamma}{m^{2}} - \frac{1}{\mu l} \right] B_{+}^{2} m_{+} \left( q_{+} \wedge \bar{D} q_{+} + m_{+} \bar{e} \wedge q_{+} \wedge q_{+} \right)\\
      + & \left[ \sigma + \frac{\alpha \beta}{\mu} + \frac{\gamma}{m^{2}} + \frac{1}{\mu l} \right] B_{-}^{2} m_{-} \left( q_{-} \wedge \bar{D} q_{-} + m_{-} \bar{e} \wedge q_{-} \wedge q_{-} \right) \\
      + & B_{1}^{2} \tilde{B}_{1} m_{1} \left( q_{1} \wedge \bar{D} q_{1} + m_{1} \bar{e} \wedge q_{1} \wedge q_{1} \right) \\
      + & B_{2}^{2} \tilde{B}_{2} m_{2} \left( q_{2} \wedge \bar{D} q_{2} + m_{1} \bar{e} \wedge q_{2} \wedge q_{2} \right)\biggr\},
  \end{split}
\end{equation}
where
\begin{equation}\label{110}
  \begin{split}
     \tilde{B}_{1} =  & - \frac{1}{2 \mu } m_{1} + \left( \sigma + \frac{\alpha \beta}{\mu} + \frac{\gamma}{m^{2}} \right) + \frac{\alpha C_{1}}{\mu} + \frac{F_{1}}{m^{2}} - \frac{1}{2 \mu l^{2} m_{1}} - \frac{\alpha ^{2} C_{1}^{2}}{2 \mu m_{1}}  \\
       & -\left( 1 + \alpha \sigma + \frac{\alpha^{2} \beta}{\mu} + \frac{\alpha\gamma}{m^{2}}\right) \frac{C_{1}}{m_{1}} - \frac{\alpha \beta F_{1}}{m^{2}m_{1}} - \frac{\alpha F_{1} C_{1}}{m^{2}m_{1}}, \\
     \tilde{B}_{2} =  & - \frac{1}{2 \mu } m_{2} + \left( \sigma + \frac{\alpha \beta}{\mu} + \frac{\gamma}{m^{2}} \right) + \frac{\alpha C_{2}}{\mu} + \frac{F_{2}}{m^{2}} - \frac{1}{2 \mu l^{2} m_{2}} - \frac{\alpha ^{2} C_{2}^{2}}{2 \mu m_{2}} \\
       & -\left( 1 + \alpha \sigma + \frac{\alpha^{2} \beta}{\mu} + \frac{\alpha\gamma}{m^{2}}\right) \frac{C_{2}}{m_{2}} - \frac{\alpha \beta F_{2}}{m^{2}m_{2}} - \frac{\alpha F_{2} C_{2}}{m^{2}m_{2}}.
  \end{split}
\end{equation}
The two first terms in the Lagrangian \eqref{109} can be written in the form of the difference of two linearized $SL(3, R)$ Chern-Simons 3-forms, so the $q_{\pm}$ fields have no local degrees of freedom for spin-2 and spin-3 fields. Then, each of two last terms in the Lagrangian \eqref{109} describes a single spin-2 massive mode and a single trace-less spin-3 massive mode. So spin-3 GMMG model propagate 2 massive graviton modes and 2 massive spin-3 modes.  The massive modes are not ghosts as long as $\tilde{B}_{1}$ and $\tilde{B}_{2}$ are both positive definite.\\
Now, we want to find the mass of spin-2 and spin-3 fields appeared in this model. To this end, using Eq.\eqref{102} and Eq.\eqref{105}, we write
\begin{equation}\label{111}
  e = \bar{e} + \hat{\varepsilon} \sum_{I} B_{I} q_{I}, \hspace{1 cm} (I=+,-,1,2).
\end{equation}
where $q_{I}$ is a $SL(3,R)$ Lie algebra valued 1-form
\begin{equation}\label{112}
  q_{I \mu}= q_{I \mu}^{\hspace{3.5 mm} a} J_{a} + q_{I \mu}^{\hspace{3.5 mm} ab} T_{ab},
\end{equation}
with $q_{I \mu a}^{\hspace{4.5 mm} a}=0$, that $q_{I \mu}^{\hspace{3.5 mm} ab}$ is trace-less in Lorentz indices. In this way, equations \eqref{108} can be written as
\begin{equation}\label{113}
  \bar{D} q_{I} + m_{I} \left( \bar{e} \wedge q_{I} +  q_{I} \wedge \bar{e} \right)=0
\end{equation}
We also mention that
\begin{equation}\label{114}
\begin{split}
     & tr\left( J_{a} J_{b}\right)=2 \eta _{ab}, \hspace{1 cm} tr \left( J_{a} T_{bc}\right)=0, \hspace{1 cm} tr\left( J_{a} J_{b} J_{c} \right)=\varepsilon_{abc}, \\
     & tr\left( J_{a} J_{b} T_{cd}\right)= - \frac{2}{3} \eta_{ab} \eta_{cd} + 2 \eta_{a(c} \eta_{d)b} , \\
     & tr\left( [J_{E},J_{A}]\{ J_{B}, J_{F} \} \right)+tr\left( [J_{E},J_{B}]\{ J_{A}, J_{F} \} \right)+tr\left( [J_{E},J_{F}]\{ J_{B}, J_{A} \} \right)=0,
\end{split}
\end{equation}
where $\eta_{ab}$ is the metric of the Minkowski space \cite{15}.\\
By substituting Eq.\eqref{111} into Eq.\eqref{2}, we will find the metric fluctuations about AdS$_{3}$ metric $\bar{g}_{\mu \nu}$ as
\begin{equation}\label{115}
  \mathbf{h}_{I \mu \nu} = \frac{1}{2} tr \left\{ \bar{e} _{\mu} q_{I\nu} \right\}= \bar{e}_{\mu a} q_{I \nu}^{\hspace{3 mm} a},
\end{equation}
where we assume that $ tr \left\{ \bar{e} _{[\mu} q_{I\nu]} \right\}=0$. Suppose that $\bar{\nabla}_{\mu}$ denotes covariant derivative with respect to the background metric $\bar{g}_{\mu \nu}$ and $\bar{\epsilon}_{\mu \nu \lambda} = \varepsilon _{abc} \bar{e}_{\mu}^{\hspace{2 mm} a} \bar{e}_{\nu}^{\hspace{2 mm} b} \bar{e}_{\lambda}^{\hspace{2 mm} c}$, also $\bar{D}^{(T)}_{\mu}$ denotes total derivative compatible with the background dreibein $\bar{e}_{\mu}^{\hspace{2 mm} A}$, i.e. $\bar{D}^{(T)}_{\mu} \bar{e}_{\nu}=0$ \footnote{$\bar{D} ^{(T)} _{\mu} \bar{e}_{\nu} = \partial _{\mu} \bar{e}_{\nu} + [ \bar{\Omega} _{\mu} , \bar{e} _{\nu}] - \bar{\Gamma} ^{\lambda} _{\mu \nu} \bar{e}_{\lambda}$.}. Therefore, we can write
\begin{equation}\label{116}
  \bar{\epsilon}^{\alpha \beta \nu } \bar{\nabla}_{\beta} \mathbf{h}_{I \mu \nu} = \frac{1}{2}\bar{\epsilon}^{\alpha \beta \nu } \bar{D}^{(T)}_{\beta} \left[ tr \left\{ \bar{e} _{\mu} q_{I\nu} \right\} \right]= \frac{1}{2}\bar{\epsilon}^{\alpha \beta \nu } tr \left\{ \bar{e} _{\mu} \bar{D}_{\beta} q_{I\nu} \right\}.
\end{equation}
By substituting Eq.\eqref{113} into Eq.\eqref{116} and by using the trace-less condition $ \mathbf{h}^{\hspace{1.5 mm} \alpha}_{I \hspace{2.5 mm} \alpha}=0$, we find the following first-order equation for $\mathbf{h}_{I}$
\begin{equation}\label{117}
  \mathcal{D}[m_{I}]^{\mu} _{\hspace{2 mm} \alpha} \mathbf{h}^{\hspace{1.5 mm} \alpha}_{I \hspace{2.5 mm} \nu}=0,
\end{equation}
where
\begin{equation}\label{118}
  \mathcal{D}[m_{I}]^{\mu} _{\hspace{2 mm} \nu} = m_{I} \delta ^{\mu} _{\hspace{2 mm} \nu} + \bar{\epsilon}^{\mu \beta }_{\hspace{3.5 mm} \nu} \bar{\nabla}_{\beta}.
\end{equation}
In order to find the second-order equation, we apply the operator $ \mathcal{D}[-m_{I}] $ on the equation \eqref{117}, that is
\begin{equation}\label{119}
  \mathcal{D}[-m_{I}]^{\mu} _{\hspace{2 mm} \alpha} \mathcal{D}[m_{I}]^{\alpha} _{\hspace{2 mm} \beta} \mathbf{h}^{\hspace{1.5 mm} \beta}_{I \hspace{2.5 mm} \nu}=0.
\end{equation}
In this way, one obtains the following second-order equation
\begin{equation}\label{120}
  \left( \bar{\Box} - \mathcal{M}_{I}^{2} + \frac{2}{l^{2}} \right) \mathbf{h}_{I \mu \nu}=0,
\end{equation}
with
\begin{equation}\label{121}
  \mathcal{M}_{I}^{2} = m_{I}^{2} - \frac{1}{l^{2}},
\end{equation}
where we have used the transverse condition $ \bar{\nabla}_{\alpha} \mathbf{h}^{\hspace{1.5 mm} \alpha}_{I \hspace{2.5 mm} \nu}=0$. It is clear that
the equation \eqref{120} is the Fierz-Pauli spin-2 field equation in AdS$_{3}$ for a spin-2 field $\mathbf{h}_{I \mu \nu}$ with mass $\mathcal{M}_{I}$ for $I=1,2$. \footnote{The massless graviton in three dimensions has no degrees of freedom, which is why people call Einstein-Hilbert three dimensional gravity a topological theory. But in the higher-derivative theories one gets in addition to the massless graviton (which is pure gauge), several massive gravitons which all have 2 degrees of freedom each.}\\
Now, we substitute Eq.\eqref{111} into Eq.\eqref{2'} then we obtain the spin-3 fluctuations as follows:
\begin{equation}\label{122}
  \mathbf{H}_{I \mu \nu \lambda} = \frac{1}{2} tr \left\{ \bar{e} _{\mu} \bar{e} _{\nu} q_{I\lambda} \right\}= \bar{e}_{\mu a} \bar{e}_{\nu b} q_{I \lambda}^{\hspace{3 mm} ab},
\end{equation}
where we assume that $ tr \left\{ \bar{e} _{[\mu} \bar{e} _{\nu} q_{I\lambda]} \right\}=0$. Also, it should be noted that $\bar{\varphi}_{\mu \nu \lambda}= \frac{1}{3!} tr(\bar{e}_{\mu} \bar{e}_{\nu} \bar{e}_{\lambda }) =0$. Therefore, we can write
\begin{equation}\label{123}
  \bar{\epsilon}^{\alpha \beta \lambda } \bar{\nabla}_{\beta} \mathbf{H}_{I \mu \nu \lambda} = \frac{1}{2}\bar{\epsilon}^{\alpha \beta \lambda } tr \left\{ \bar{e} _{\mu} \bar{e} _{\nu} \bar{D}_{\beta} q_{I\lambda} \right\}.
\end{equation}
By substituting Eq.\eqref{113} into Eq.\eqref{123}, we find the following first-order equation for $\mathbf{H}_{I}$
\begin{equation}\label{124}
  \tilde{\mathcal{D}}[m_{I}]^{\mu} _{\hspace{2 mm} \alpha} \mathbf{H}^{\hspace{1.5 mm} \alpha}_{I \hspace{2.5 mm} \nu \lambda}=0,
\end{equation}
with
\begin{equation}\label{125}
  \tilde{\mathcal{D}}[m_{I}]^{\mu} _{\hspace{2 mm} \nu} = 2 m_{I} \delta ^{\mu} _{\hspace{2 mm} \nu} + \bar{\epsilon}^{\mu \beta }_{\hspace{3.5 mm} \nu} \bar{\nabla}_{\beta}.
\end{equation}
where we have used the trace-less condition $ \mathbf{H}^{\hspace{1.5 mm} \alpha}_{I \hspace{2.5 mm} \alpha \mu \nu}=0$. Now, we apply the operator $ \tilde{\mathcal{D}}[-m_{I}] $ on the equation \eqref{124}, that is
\begin{equation}\label{126}
  \tilde{\mathcal{D}}[-m_{I}]^{\mu} _{\hspace{2 mm} \alpha} \tilde{\mathcal{D}}[m_{I}]^{\alpha} _{\hspace{2 mm} \beta} \mathbf{H}^{\hspace{1.5 mm} \beta}_{I \hspace{2.5 mm} \nu \lambda}=0.
\end{equation}
Then, one obtains the following second-order equation
\begin{equation}\label{127}
  \left( \bar{\Box} - \tilde{\mathcal{M}}_{I}^{2} \right) \mathbf{H}_{I \mu \nu \lambda}=0,
\end{equation}
with
\begin{equation}\label{128}
  \tilde{\mathcal{M}}_{I}^{2} = 4 \mathcal{M}_{I}^{2} = 4 \left( m_{I}^{2} - \frac{1}{l^{2}}\right),
\end{equation}
where we have used the transverse condition $ \bar{\nabla}_{\alpha} \mathbf{H}^{\hspace{1.5 mm} \alpha}_{I \hspace{2.5 mm} \mu \nu}=0$. The equation \eqref{127} is the spin-3 field equation in AdS$_{3}$ for a spin-3 field $\mathbf{H}_{I \mu \nu \lambda}$ with mass $\tilde{\mathcal{M}}_{I}$. It is clear from Eq.\eqref{121} and Eq.\eqref{128} that "no-tachyon" condition can be written as $\left| l m_{I} \right| \geq 1$. Because the $q_{\pm}$ fields have no local degrees of freedom, then $\mathbf{H}_{+ \mu \nu \lambda}$ and $\mathbf{H}_{- \mu \nu \lambda}$ are not propagating modes. Thus, we have two massive propagating modes $\mathbf{H}_{1 \mu \nu \lambda}$ and $\mathbf{H}_{2 \mu \nu \lambda}$.
\section{Example}
In this section we consider the Spin-3 GMMG, then we find energy, angular momentum and entropy of a special black hole solution. In higher spin theories the metric is gauge dependent, so may be one say that the notion of an event horizon for a black hole is not well defined concept \cite{105}. However different black hole solutions in higher spin gravity have been obtained \cite{14,105,106,107,31,2,108,32,109,110,111,15,112,101,102}. In some of these papers also different methods for deriving  entropy of higher spin black holes have been presented \cite{14,112,2,32}. In spin-3 gravity in contrast to the ordinary gravity, where entropy of black hole is given by the area of the horizon, since the metric is gauge dependent, this statement should be replaced with a gauge invariant criterion \cite{105}. Until now there is not a general formula for entropy of all black hole solutions in higher spin gravity models. Also there are some discrepancy between assumptions and results of different methods which have been presented for deriving the entropy of higher spin black holes. Here we present a formula for obtaining entropy of higher spin black holes. \\
In order to relate Chern-Simons gauge theory \eqref{280} to the first order formalism based on dreibein and spin-connection we introduce following two independent connection one-form $A^{+}$ and $A^{-}$,
\begin{equation}\label{300}
  A ^{\pm} = \omega \pm \frac{1}{l} e .
\end{equation}
Now we consider following gauge connections which solve the spin-3 gravity field equations \cite{1,14}
\begin{equation}\label{87}
  \begin{split}
       & A^{+} = b(\rho)^{-1} a^{+}(x^{+}) b(\rho)+b(\rho)^{-1} d b(\rho) \\
       & A^{-} = b(\rho) a^{-}(x^{-}) b(\rho)^{-1}+b(\rho) d b(\rho)^{-1},
  \end{split}
\end{equation}
where $b(\rho)= \exp{(\rho L_{0})}$ and $x^{\pm}= t/l \pm \phi$. Also, $a^{\pm}(x^{\pm})$ are given by
\begin{equation}\label{88}
  \begin{split}
       & a^{+}(x^{+})= \left( L_{1} - \mathcal{L}^{+}(x^{+}) L_{-1} - \mathcal{W}^{+}(x^{+}) W_{-2} \right) dx^{+} \\
       & a^{-}(x^{-})= \left( L_{-1} - \mathcal{L}^{-}(x^{-}) L_{1} + \mathcal{W}^{-}(x^{-}) W_{2} \right) dx^{-}.
  \end{split}
\end{equation}
where $\mathcal{L}^{\pm}(x^{\pm})$ and $\mathcal{W}^{\pm}(x^{\pm})$ are functions which transform under gauge transformations which preserve the asymptotic conditions (see \cite{1} for details). By substituting the gauge connections \eqref{87} into Eq.\eqref{300} we can find the generalized dreibein as
\begin{equation}\label{89}
  \begin{split}
       & e_{t}=\frac{1}{2} \left( (e^{\rho} - \mathcal{L}^{-} e^{-\rho} ) L_{1} + ( e^{\rho} - \mathcal{L}^{+} e^{-\rho} ) L_{-1} + \mathcal{W}^{-} e^{-2\rho} W_{2} - \mathcal{W}^{+} e^{-2\rho} W_{-2} \right) \\
       & e_{\phi}=\frac{l}{2} \left( (e^{\rho} + \mathcal{L}^{-} e^{-\rho} ) L_{1} - ( e^{\rho} + \mathcal{L}^{+} e^{-\rho} ) L_{-1} - \mathcal{W}^{-} e^{-2\rho} W_{2} - \mathcal{W}^{+} e^{-2\rho} W_{-2} \right) \\
       & e_{\rho}= l L_{0}.
  \end{split}
\end{equation}
Similarly, the space-time components of the generalized spin-connections can be find as
\begin{equation}\label{90}
  \omega _{t} = \frac{1}{l^2} e_{\phi}, \hspace{0.7 cm} \omega _{\rho} =0, \hspace{0.7 cm} \omega _{\phi} = e _{t}.
\end{equation}
By substituting generalized dreibeins Eq.\eqref{89} into Eq.\eqref{2}, the metric takes following form
\begin{equation}\label{91}
  \begin{split}
     ds^{2} = & - \left( \mathcal{L}^{+} \mathcal{L}^{-} e^{-2 \rho} + 4 \mathcal{W}^{+} \mathcal{W}^{-} e^{-4 \rho} - \mathcal{L}^{+} - \mathcal{L}^{-} +e^{2 \rho} \right) dt^{2} \\
     & + l^{2} d \rho ^{2} + 2 l (\mathcal{L}^{+} - \mathcal{L}^{-}) dt d \phi \\
       & + l^{2} \left( \mathcal{L}^{+} \mathcal{L}^{-} e^{-2 \rho} + 4 \mathcal{W}^{+} \mathcal{W}^{-} e^{-4 \rho} + \mathcal{L}^{+} + \mathcal{L}^{-} +e^{2 \rho} \right) d \phi ^{2}.
  \end{split}
\end{equation}
One can rewrite the above metric as the following form
\begin{equation}\label{980}
  ds^{2}=g_{tt}dt^{2}+g_{\rho \rho} d \rho ^{2} +g_{\phi \phi} d \phi ^{2} +2 g_{t \phi} dt d \phi.
\end{equation}
We have $g_{t \phi} ^{2}- g_{tt} g_{\phi \phi}=0$ on the Killing horizon $H$ which leads to the following equation
\begin{equation}\label{990}
  \mathcal{L}^{+} \mathcal{L}^{-} e^{-2 \rho _{H}} + 4 \mathcal{W}^{+} \mathcal{W}^{-} e^{-4 \rho _{H}} + e^{2 \rho _{H}}= 2 \sqrt{\mathcal{L}^{+} \mathcal{L}^{-}}.
\end{equation}
where we suppose that the Killing horizon is located at $\rho = \rho _{H}$. In previous paper \cite{4} we have shown that $\rho _{H}$ is a real positive-definite root of Eq.\eqref{990}. However we are not sure that the metric \eqref{91} for non-zero $\mathcal{W}^{\pm}$ can describe a black hole solution of higher-spin gravity. But if we extend the $a^{\pm}(x^{\pm})$ in Eq.\eqref{88} to contain the chemical potential conjugate to the $W$ charges, then we have a black hole with spin-3 charge.\footnote{The classical solutions of higher spin generalization of TMG, include AdS pp-wave (with higher
spin hair), the spacelike, timelike, null warped $AdS_3$
spacetimes and also spacelike warped $AdS_3$ black hole have been obtained in \cite{101}. Also the higher spin black holes in a truncated version of higher spin gravity
in $AdS_3$ have been studied in \cite{102}.}
Also we should note that an important property of higher-spin gravity is that the
  gauge transformation of higher-spin field act nontrivially on the metric. Due to this feature of the theory, the event horizon of black hole become gauge dependent \cite{16}. The authors of \cite{16} have discussed on the solutions of spin-3 gravity previously introduced in \cite{14}, and have shown that those solutions describe a traversable wormhole connecting two asymptotic region, instead a black hole. Then they have shown that under a higher spin gauge transformation these solutions can be transformed
to describe black holes with manifestly smooth event horizons.\\
The generalized off-shell ADT conserved charge of the Spin-3 GMMG can be obtained as
\begin{equation}\label{62}
\begin{split}
   \tilde{\mathcal{Q}} _{ADT} (a, \delta a ; \xi)= & tr \{ - \left( \sigma i_{\xi} e + \frac{1}{m^{2}} i_{\xi} f + \frac{\alpha}{\mu} i_{\xi} h - \frac{1}{\mu} (i_{\xi} \Omega - \lambda _{\xi})  \right) \delta \Omega \\
     &  - (i_{\xi} \Omega - \lambda _{\xi}) \left( \sigma \delta e + \frac{1}{m^{2}} \delta f + \frac{\alpha}{\mu} \delta h \right) + \frac{\alpha ^{2}}{\mu} i_{\xi} h \delta h \\
     &  + (1 + \alpha \sigma) ( i_{\xi} e \delta h + i_{\xi} h \delta e ) + \frac{\alpha}{m^{2}} (i_{\xi} f \delta h + i_{\xi} h \delta f )  \},
\end{split}
\end{equation}
when $h$ and $f$ is given by ansatz \eqref{57}, this equation reduces to the following
\begin{equation}\label{63}
\begin{split}
   \tilde{\mathcal{Q}} _{ADT} (a, \delta a ; \xi)= & - tr \{  \left( \left( \sigma + \frac{\gamma}{m^{2}} + \frac{\alpha \beta}{\mu} \right) i_{\xi} e - \frac{1}{\mu} ( i_{\xi} \Omega - \lambda _{\xi} )  \right) \delta \Omega \\
     &  +  \left( \left( \sigma + \frac{\gamma}{m^{2}} + \frac{\alpha \beta}{\mu} \right) (i_{\xi} \Omega - \lambda _{\xi}) - \frac{1}{\mu l ^{2}} i_{\xi} e  \right) \delta e \}.
\end{split}
\end{equation}
Now, we take AdS$_{3}$ spacetime as background solution, where
\begin{equation}\label{64}
\begin{split}
     & e _{t}=\frac{1}{2} e^{\rho} \left( L_{1} + L_{-1}  \right),
      \hspace{0.5 cm} e _{\phi}=\frac{l}{2} e^{\rho} \left(  L_{1} -  L_{-1} \right)
      , \hspace{0.5 cm} e _{\rho}= l L_{0} \\
     & \Omega _{t} = \frac{1}{l^2} e_{\phi}, \hspace{0.7 cm} \Omega _{\rho} =0, \hspace{0.7 cm} \Omega _{\phi} = e _{t}
\end{split}
\end{equation}
and its perturbation given by \cite{1}
\begin{equation}\label{65}
  \begin{split}
       & \delta e_{t}= - \frac{1}{2} e^{-\rho} \left( \delta \mathcal{L}^{-} L_{1} + \delta \mathcal{L}^{+} L_{-1} + e^{-\rho} \delta \mathcal{W}^{-} W_{2} + e^{-\rho} \mathcal{W}^{+} W_{-2} \right), \\
       & \delta e_{\phi}= - \frac{l}{2} e^{-\rho} \left( - \delta \mathcal{L}^{-} L_{1} + \delta \mathcal{L}^{+} L_{-1} - e^{-\rho} \delta \mathcal{W}^{-} W_{2} + e^{-\rho} \delta \mathcal{W}^{+} W_{-2} \right), \\
       & \delta e_{\rho}= 0 ,
  \end{split}
\end{equation}
Energy corresponds to the Killing vector $\xi_{(t)}= - \partial _{t} $, so we have
\begin{equation}\label{66}
  \lambda _{\xi _{(t)}} - i _{\xi _{(t)}} \Omega = \frac{1}{l^{2}} e _{\phi}.
\end{equation}
 Therefore, Eq.\eqref{63} becomes
\begin{equation}\label{67}
  \tilde{\mathcal{Q}} _{ADT} (- \partial _{t}) = 2 \left[ \left( \sigma + \frac{\gamma}{m^{2}} + \frac{\alpha \beta}{\mu}\right)
   ( \delta \mathcal{L}^{+} + \delta \mathcal{L}^{-} ) +  \frac{1}{\mu l} (\delta \mathcal{L}^{-} - \delta \mathcal{L}^{+} ) \right].
\end{equation}
We take $\Sigma$ as a circle of arbitrary radii, then by substituting Eq.\eqref{67} into Eq.\eqref{43} we find the energy of the considered solution as
\begin{equation}\label{68}
  E = \frac{1}{4 \pi}  \int_{0}^{2 \pi} d \phi \left[ \left( \sigma + \frac{\gamma}{m^{2}} + \frac{\alpha \beta}{\mu}\right)
   ( \mathcal{L}^{+} + \mathcal{L}^{-} ) +  \frac{1}{\mu l} ( \mathcal{L}^{-} - \mathcal{L}^{+} ) \right].
\end{equation}
Angular momentum corresponds to the Killing vector $\xi_{(\phi)}= \partial _{\phi} $ then we have
\begin{equation}\label{69}
  \lambda _{\xi _{(\phi)}} - i _{\xi _{(\phi)}} \Omega = e _{t}.
\end{equation}
 Therefore, for this case, Eq.\eqref{63} becomes
\begin{equation}\label{70}
  \tilde{\mathcal{Q}} _{ADT} ( \partial _{\phi}) = 2 l \left[ \left( \sigma + \frac{\gamma}{m^{2}} + \frac{\alpha \beta}{\mu}\right)
   ( \delta \mathcal{L}^{-} - \delta \mathcal{L}^{+} ) +  \frac{1}{\mu l} (\delta \mathcal{L}^{+} + \delta \mathcal{L}^{-} ) \right],
\end{equation}
 by substituting Eq.\eqref{70} into Eq.\eqref{43} we find the angular momentum of the considered solution as
\begin{equation}\label{71}
  j = \frac{l}{4 \pi}  \int_{0}^{2 \pi} d \phi \left[ \left( \sigma + \frac{\gamma}{m^{2}} + \frac{\alpha \beta}{\mu}\right)
   ( \mathcal{L}^{-} - \mathcal{L}^{+} ) +  \frac{1}{\mu l} ( \mathcal{L}^{+} + \mathcal{L}^{-} ) \right].
\end{equation}
Now, we apply the formula \eqref{49} to calculate the entropy of the considered black hole solution. Since the components of the flavour metric is given by Eq.\eqref{50} and $\Omega = \omega -\alpha h$, then
\begin{equation}\label{72}
   S = \int_{\mathcal{B}}  \frac{d \phi}{\sqrt{ g_{\phi \phi}}} tr \{ \sigma e_{\phi} e_{\phi} + \frac{1}{m^{2}} e_{\phi} f _{\phi} - \frac{1}{\mu} e_{\phi} \Omega _{\phi} + \frac{\alpha}{\mu} e_{\phi} h_{\phi} \} .
\end{equation}
By substituting $f$ and $h$ from Eq.\eqref{57} into above equation we find that
\begin{equation}\label{73}
   S = 2 \int_{0}^{2 \pi}  \frac{d \phi}{\sqrt{ g_{\phi \phi}}}  \left[ \left( \sigma + \frac{\gamma}{m^{2}} + \frac{\alpha \beta }{\mu} \right) g_{\phi \phi} - \frac{1}{\mu} g_{t \phi} \right] .
\end{equation}
where we used Eq.\eqref{2} in the last step of calculation. If we set $e_{\mu} ^{\hspace{1.5 mm} ab}= \omega_{\mu} ^{\hspace{1.5 mm} ab}=0 $, then the above formula reduces to the entropy formula in the ordinary GMMG which is obtained in the paper \cite{90}.
One can substitutes $g_{\phi \phi}$ and $g_{t \phi}$ from \cite{4} then
\begin{equation}\label{74}
   S = 2 \int_{0}^{2 \pi}  \left[ \left( \sigma + \frac{\gamma}{m^{2}} + \frac{\alpha \beta }{\mu} \right) (\sqrt{\mathcal{L}^{-}} + \sqrt{\mathcal{L}^{+}}) - \frac{1}{\mu} (\sqrt{\mathcal{L}^{-}} - \sqrt{\mathcal{L}^{+}}) \right] d \phi.
\end{equation}
If we set $\alpha =0$ and in the limit $m^{2} \rightarrow \infty$, where Spin-3 GMMG model reduce to the spin-3 TMG model, the above equations for energy, angular momentum and entropy in Eqs. \eqref{68}, \eqref{71} and \eqref{74} reduce to the corresponding results in the spin-3 TMG \cite{4}. Also as we have shown in \cite{4}, in the limiting case $\frac{1}{\mu}\rightarrow 0$, where TMG action reduce to the Einstein-Hilbert action, our conserved charge in this limiting case reduce to the result presented in papers \cite{31,32}. This coincidence of results confirm that the quasi-local formalism works correctly for these types of theories.\\
Now, we want to compare Eq.\eqref{74} with the obtained result for entropy of the BTZ black hole \cite{103} solution of ordinary MMG (see Ref.\cite{104}). The ordinary MMG can be seen as a limiting case of ordinary GMMG model. To find corresponding results in ordinary MMG model, we must set $e_{\mu} ^{\hspace{1.5 mm} ab}= \omega_{\mu} ^{\hspace{1.5 mm} ab}=0 $ and $m^{2} \rightarrow \infty$. In this limit Eq.\eqref{59} and Eq.\eqref{60} become
\begin{equation}\label{129}
  \Lambda _{0} - \alpha (1+ \alpha \sigma) \beta ^{2} = -\frac{\sigma}{l^{2}},
\end{equation}
\begin{equation}\label{130}
  \alpha ^{2} \beta ^{2} + 2 \mu (1+\alpha \sigma) \beta = \frac{1}{l^{2}}.
\end{equation}
In this case we do not consider Eq.\eqref{61} because Eq.\eqref{61} comes from the variation of Lagrangian of spin-3 GMMG with respect to auxiliary field $f$ and we lose $f$ as $m^{2}$ tends to infinity (see Eq.\eqref{100}). From equations \eqref{129} and \eqref{130}, one finds
\begin{equation}\label{131}
  \beta=\frac{1-\alpha \Lambda_{0}l^{2}}{2 \mu l^{2} \left( 1+ \alpha \sigma\right)^{2}}.
\end{equation}
For the BTZ black hole we have
\begin{equation}\label{132}
  \mathcal{L}^{\pm}= \left( \frac{ r_{+} \pm r_{-}}{2l}\right)^{2},
\end{equation}
where $r_{+}$ and $r_{-}$ are outer and inner horizon radiuses, respectively. By substituting Eq.\eqref{131} and \eqref{132} into the Eq.\eqref{74} and taking $m^{2} \rightarrow \infty$ we will obtain entropy of the BTZ black hole solution of ordinary MMG which has been calculated in \cite{104}. The same arguments are hold for energy and angular momentum of the BTZ black hole.
\section{Conclusion}
In this paper, we considered the spin-3 gravity in the first order formalism, then we generalized that to the Spin-3 CSLTG. It should be noted that the Spin-3 TMG \cite{9,10} is an example of such theories. We provided a general formula to compute the conserved charges and the entropy for solutions in these theories that are generalizations of the standard three-dimensional higher spin gravity.\\ We found the off-shell ADT current \eqref{38} associated to a vector field $\xi$ for the Spin-3 CSLTG. This current is conserved when $\xi$ is a Killing vector field. We defined the off-shell generalized ADT current by \eqref{40}, so that it becomes conserved for an asymptotically Killing vector field as well as a Killing vector field admitted by the whole of spacetime. The generalized off-sell ADT current reduces to the ordinary one when $\xi$ is a Killing vector field which admitted by the whole of spacetime and it reduces to the simplectic current when equations of motion and linearized equations of motion are satisfied. We defined the generalized off-shell ADT conserved charge \eqref{42} through the generalized off-sell ADT current. We used the generalized off-shell ADT conserved charge in-order to define quasi-local conserved charge Eq.\eqref{44}. The obtained quasi-local conserved charge \eqref{44} is associated to an asymptotically Killing vector field $\xi$ and the integration surface $\Sigma$ can be choose arbitrarily. The advantage of this formalism to find conserved charge is that it is applicable to solutions that are not asymptotically (A)dS. In section 5, we found a general formula for entropy of black holes in the context of Spin-3 CSLTG \eqref{49}. In section 6, we considered the Spin-3 GMMG as an example of the Spin-3 CSLTG and we have obtained a class of solutions for that model. We have found the quadratic Lagrangian for the fluctuations $u$, $v$, $w$ and $z$. Then by introduction a transformations from $(u,v,w,z)$ to new Lie algebra valued one-form fluctuations $(q_{+},q_{-},q_{1},q_{2})$, we could obtain diagonalize form of mentioned Lagrangian. The two first terms in the Lagrangian Eq.\eqref{109} can be written in the form of the difference of two linearized $SL(3, R)$ Chern-Simons 3-forms, so the $q_{\pm}$ fields have no local degrees of freedom for spin-2 and spin-3 fields, as we expect. Then, each of two last terms in the Lagrangian  Eq.\eqref{109} describes a single spin-2 massive mode and a single trace-less spin-3 massive mode. So spin-3 GMMG model propagate 2 massive graviton modes and 2 massive spin-3 modes.
 The massive modes are not ghosts as long as $\tilde{B}_{1}$ and $\tilde{B}_{2}$ are both positive definite.  We have shown that the spin-2 fluctuations $\mathbf{h}_{I \mu \nu}$ satisfy the Fierz-Pauli spin-2 field equation \eqref{120} in AdS$_{3}$ space with mass $\mathcal{M}_{I}$ and, the spin-3 fluctuations $\mathbf{H}_{I \mu \nu \lambda}$ satisfy the spin-3 field equation \eqref{127} in AdS$_{3}$ space with mass $\tilde{\mathcal{M}}_{I}$. Also, we deduced that no-tachyon condition is $\left| l m_{I} \right| \geq 1$. Eventually, in section 7, we obtained energy, angular momentum and entropy of a special black hole solution in the context of the Spin-3 CSLTG.\\
In subsection \ref{S.A}, we analysis spin-3 fluctuation. We focused on its traceless part. There is actually a trace part of spin-3 fluctuations. Such a problem has been studied in the context of spin-3 TMG \cite{10}. The Lagrangian of spin-3 GMMG will reduce to spin-3 TMG one when we set $m^{2} \rightarrow \infty$ and $\alpha =0$. In that case the quadratic Lagrangian for the fluctuations \eqref{109} just contains one massive mode Lagrangian (i.e. in the spin-3 TMG limit, one of the coefficients $\tilde{B} _{1}$ or $\tilde{B} _{2}$ will vanish.). In the spin-3 TMG model, massive trace mode has zero energy and becomes pure gauge at the chiral point. In a similar way, in the context of spin-3 GMMG, we expect that massive trace modes become pure gauges at the chiral point $\sigma + \frac{\alpha \beta}{\mu} + \frac{\gamma}{m^{2}} = \frac{1}{\mu l}$ when two operators $\mathcal{D}[m_{1}]$ and $\mathcal{D}[m_{2}]$ are degenerate (namely, when $m_{1} = m_{2}$). Therefore, our conclusions are unaffected for chiral spin-3 GMMG with $m_{1} = m_{2}$.\\
It would be interesting to study a Hamiltonian analysis of Spin-3 CSLTG as the authors of \cite{5} have done for ordinary CSLTG. For example a Hamiltonian analysis  show that the GMMG model has
no Boulware-Deser ghosts and this model propagate only two physical modes \cite{8}. GMMG also avoids the aforementioned ``bulk-boundary unitarity clash''. So it is a semi-classical quantum gravity model in $2+1$ dimension which in both bulk and boundary is unitary. Therefore it is clear that study of Spin-3 GMMG following what have done in the papers \cite{8,9,10} is interesting request. We let this study for future work.
\section{Acknowledgments}
M. R. Setare thank  Wontae Kim, and Alfredo Perez  for reading the manuscript, helpful comments and discussions.
\appendix

\section{$ SL(3,\mathbb{R})$ generators}
We use the following basis of $ SL(3,\mathbb{R})$ generators \cite{1,14}
\begin{equation}\label{A1}
  \begin{split}
       & L_{1}=\left(
                 \begin{array}{ccc}
                   0 & 0 & 0 \\
                   1 & 0 & 0 \\
                   0 & 1 & 0 \\
                 \end{array}
               \right)
        \hspace{0.5 cm}
        L_{0}=\left(
          \begin{array}{ccc}
            1 & 0 & 0 \\
            0 & 0 & 0 \\
            0 & 0 & -1 \\
          \end{array}
        \right)
        \hspace{0.5 cm}
        L_{-1}=\left(
          \begin{array}{ccc}
            0 & -2 & 0 \\
            0 & 0 & -2 \\
            0 & 0 & 0 \\
          \end{array}
        \right)
        \\
       &W_{2}=\left(
          \begin{array}{ccc}
            0 & 0 & 0 \\
            0 & 0 & 0 \\
            2 & 0 & 0 \\
          \end{array}
        \right)
         \hspace{0.5 cm}
         W_{1}=\left(
          \begin{array}{ccc}
            0 & 0 & 0 \\
            1 & 0 & 0 \\
            0 & -1 & 0 \\
          \end{array}
        \right)
        \hspace{0.5 cm}
         W_{0}=\frac{2}{3}\left(
          \begin{array}{ccc}
            1 & 0 & 0 \\
            0 & -2 & 0 \\
            0 & 0 & 1 \\
          \end{array}
        \right)
        \\
       & W_{-1}=\left(
          \begin{array}{ccc}
            0 & -2 & 0 \\
            0 & 0 & 2 \\
            0 & 0 & 0 \\
          \end{array}
        \right)
        \hspace{0.5 cm}
         W_{-2}=\left(
          \begin{array}{ccc}
            0 & 0 & 8 \\
            0 & 0 & 0 \\
            0 & 0 & 0 \\
          \end{array}
        \right).
  \end{split}
\end{equation}
In other words, we take $ J_{A}= \{ L_{0}, L_{\pm 1}, W_{0},W_{\pm 1}, W_{\pm 2} \} $.The generators obey the following commutation relations
\begin{equation}\label{A2}
  \begin{split}
       & [L_{i},L_{j}]=(i-j) L_{i+j} \\
       & [L_{i},W_{m}]=(2i-m) W_{i+m} \\
       & [W_{m},W_{n}]=-\frac{1}{3}(m-n)(2m^{2}+2n^{2}-mn-8)L_{m+n}.
  \end{split}
\end{equation}
where $-1 \leq i,j \leq 1 $ and $-2 \leq m,n \leq 2$. Also, the non-zero traces are
\begin{equation}\label{A3}
  \begin{split}
       & tr(L_{0}L_{0})=2, \hspace{0.5 cm} tr(L_{1}L_{-1})=-4 \\
       & tr(W_{0}W_{0})=\frac{8}{3}, \hspace{0.5 cm} tr(W_{1}W_{-1})=-4,\hspace{0.5 cm} tr(W_{2}W_{-2})=16 .
  \end{split}
\end{equation}
The Killing form in the fundamental representation of $sl(3, \mathbb{R})$ is defined as
\begin{equation}\label{A4}
  K_{AB}=\frac{1}{2} tr(J_{A} J_{B}),
\end{equation}
and anti-symmetric and symmetric structure constants of the Lie algebra are given by
\begin{equation}\label{A5}
  f_{ABC}=\frac{1}{2} tr([ J_{A} , J_{B} ]J_{C}), \hspace{1 cm} d_{ABC}=\frac{1}{2} tr(\{ J_{A} , J_{B} \} J_{C}) .
\end{equation}

\end{document}